\documentclass[11pt,twoside]{article}

\usepackage{asp2006}

\begin{document}
\title{Fourier Disentangling Using the Technology of Virtual Observatory}   
\author{Petr \v{S}koda and Petr Hadrava  }   
\affil{Astronomical Institute Academy of Sciences, Fri\v{c}ova 298, 
 251 65 Ond\v{r}ejov, Czech Republic}    

\begin{abstract} 
The Virtual Observatory is a new technology of the  astronomical research
allowing the  seamless processing and analysis of a  heterogeneous data
obtained from a number of distributed data archives.  It may also provide
astronomical community with powerful computational and data processing on-line
services replacing the custom scientific code run on user's computers.

Despite its benefits the VO technology has been still  little exploited in
stellar spectroscopy. As an example of possible evolution in this field we
present an experimental web-based  service for disentangling of spectra based on
code KOREL.  This code developed by P.  Hadrava enables Fourier disentangling
and line-strength photometry, i.e. simultaneous decomposition of spectra of
multiple stars and solving for  orbital parameters, line-profile variability or
other physical parameters of observed objects.

We discuss the benefits of the service-oriented approach from the point of view
of both developers and users and give examples of possible user-friendly
implementation of spectra disentangling methods as a standard tools of Virtual
Observatory.
\end{abstract}
%
%
\section{Introduction}   
The astronomical spectroscopy uses many special techniques to analyse
stellar spectra and estimate physical properties of targets studied.
Basically they consist in comparison of the observed spectra with
theoretical models which, however, may be of very different level
of sophistication. For instance, a simple comparison of suitably defined
effective centres of spectral lines with their laboratory wavelengths
gives Doppler shifts, which in the case of spectroscopic binaries
enables one to determine their orbital parameters. Detailed comparison
of equivalent widths and shapes of line profiles with synthetic spectra
may reveal effective temperatures, gravity acceleration, abundances and
other physical parameters of stellar atmospheres. In practice, however,
the spectra of components of the binary are blended and the information
on orbital and atmospheric parameters are entangled.

 Several techniques for separation of component spectra from a series
of spectra has been proposed which enable also to develop the so called
spectra disentangling, i.e. a method of simultaneous separation of the
spectra and determination of physical parameters governing their
variability. In particular, the method of Fourier disentangling introduced
and implemented in program KOREL by \citet{h95} proved to be efficient
and viable for a further generalisation.

To allow the application of such a powerful method on a number of different
objects in a scalable way, we attempted to embed  the KOREL in the
infrastructure of Virtual Observatory.
\section{The Virtual Observatory}
Contemporary astronomy faces an enormous amount of data continuously flowing
from large telescopes, space missions and supercomputer simulations, that can
hardly be analysed (and even previewed) by the traditional scientific methods.
Thus the concept of (astronomical) Virtual Observatory (VO) was recently born
aiming at federalisation of all  astronomical resources (e.g. catalogues, data
archives, simulation databases, data processing and analysing tools) using  the
global infrastructure based on  unified data format  and  set of rigid,  yet
extensible  communication protocols.
The development and implementation of these global standards is the role of the
International Virtual Observatory Alliance (IVOA).

Technically, VO is a collection of inter-operating data archives and software
tools which utilise the internet to form a  virtual desktop environment in
which astronomical research  can be conducted in a user friendly manner
allowing the astronomer to concentrate on asking the scientific questions
instead of spending most of the  time with  searching  in  heterogeneous
scattered archives, and with  homogenisation of data  represented by  different
units in various file formats.

Owing to its  huge data-mining potential and easy multiwavelength analysis
tools, the VO technology allows  to tackle problems not feasible by any other
means, like the search of very rare astronomical events,  candidates of yet
unknown classes of objects (e.g. extremely cold brown dwarfs, supermassive
stars etc.), statistics   of order of tens of  millions target or pan-spectral
classification as building the spectra energy distributions of  radiation from
gamma to radio using the archives of all space and ground-based observations
together. For the extensive introduction into the VO science see
\citet{2006LNEA....2...71S}.
\section{The Fourier  Disentangling}
The disentangling of spectra represents nowadays a whole branch of stellar
spectroscopy fairly exceeding the scope of 
our contribution. We thus refer for a detailed
explanation of its physical and mathematical principles, astrophysical
consequences and for corresponding literature to the review \citep*{h04} or its
update \citep*{h09b}. Here we shall only qualitatively characterise the method
of Fourier disentangling implemented in code KOREL and we shall list a recent progress.

 The instantaneous spectra of many variable objects can be in a good
 approximation expressed as a superposition of their intrinsic (time
 independent) components convolved with some broadening functions (e.g.
 Doppler shifted delta-functions) depending on time and some physical
 parameters of the variability (e.g. the orbital parameters). In the Fourier
 conjugate space the intrinsic components can thus easily be solved
 (independently for each Fourier mode) from a more numerous set of
 observations. Moreover, the values of the free parameters can be fit by the
 least-square method. To prevent an ill-determination of the problem,
 a  good coverage of the  time interval of the characteristic variability is
 needed. The main task of the development of the method is thus to find a
 proper theoretical model of the broadening.  Already the very simple
 assumption of line-strength variability with fixed line profiles \citep*{h97}
 enables many useful applications.  To apply the method successfully to real
 data, the observers should understand the assumptions and properties of the
 model and to prepare a set of data decisive for the parameters required from
 the solution.

 If the solution for the intrinsic component spectra is well over-determined by
 a great number of observed spectra, their noise can be substantially reduced
 by the averaging. A recent improvement of the numerical technique
 \citep*{h09a} enables to retrieve the radial-velocity shifts with an accuracy
 surpassing the limitations by the step of spectra sampling (this is sometimes
 called super-resolution). Our recent work \citep*{hss09} opens a disentangling
 of Cepheid pulsations.
\section{The Virtual Observatory Web Services}
As the Fourier disentangling of the large number of spectra may become computation
intensive, its full power may be exploited  using the modern technology of VO Web
Services (WS). 
The WS is typically complex processing application using
the web technology (http protocol and (X)HTML markup) to transfer input data
(files, tables, images, spectra etc.) to the main processing back-end (often
run in front of queue scheduling and/or parallelising engine on computer
clusters or GRIDS) and the results (after intensive number crunching) back to
user.  All this can be done using only an ordinary web browser (and in
principle the science may be done on the fast palmtop or advanced mobile
phone). 

The more detailed analysis about the benefits of GRID technology in
stellar spectroscopy is presented by \citet{2009MmSAI..80..484S}.
This service-oriented approach has many advantages both for the user and
developer. Let's name some of them:
\begin{itemize} 
\item There is the only one, current, well tested version of
the code (and documentation), maintained and updated by its author 
\item The user needs not to install anything from the author 
\item The code is optimised for given HW (native compiler), knowing its limits
(memory and cache sizes, number of nodes etc.) 
\item The problem is scalable - the more user requirements may
be solved by adding more computing nodes and introducing priority queues 
\item The web technology provides the easy way of interaction (forms) and graphics
output (in-line images) even produced dynamically (variable refresh rates or
event driven - e.g. AJAX) 
\end{itemize}
\subsection{The KOREL Web Service}
The idea of our service is to  have an user interface similar to e-shop portal,
starting with user registration. Every set of input parameters creates a job,
which may be run in parallel with others, the user may stop or remove them, can
return to the previous versions etc.  Privileged users may even recompile their
own version of KOREL code tailored to their needs (e.g. maximum amount and size
of spectra).  All user communication is encrypted and the user can see only
his/her jobs.  The service may be accessed from the KOREL portal at
Astronomical Institute in
Ond\v{r}ejov\footnote{http://stelweb.asu.cas.cz/vo-korel}.

At the time of  preparation of the proceedings the KOREL Web Service requires
to upload the files {\tt korel.data} and {\tt korel.par} in given strict
format. Usually, for the preparation of input data the program {\tt PREKOR}
run at local computer  is used, which reads spectra in various formats, rebins
them equidistant in radial velocity (logarithmic wavelength) and optionally
applies the precisely computed heliocentric correction. In addition to that,
its interactive graphics helps to select the proper spectral regions bordered
by the clear continuum and allows the removal of bad spectra.  

In the future, the role of the {\tt PREKOR} may be replaced by another set of
web services acquiring the spectra directly from VO servers and using proper
metadata (e.g. elements of orbits) obtained from  proper catalogues published
in VO (especially CDS Vizier and Simbad). The interactive capability will be
provided by VO spectral tools (e.g. SPLAT or VOSpec).
\section{Conclusions} 
The Fourier disentangling is already well-established method of stellar spectra
analysis with the wide range of applications.  The KOREL web service is
probably one of the first attempts to adapt the legacy stellar spectra analysis
code for the Virtual Observatory service.  The advantages of solution adopted
are evident, although some level of user conservatism has to be expected.
%
%
\acknowledgements 
This work was supported by grants GA\v{C}R 202/06/0041, GA\v{C}R 202/09/0772 and by projects
AV0Z10030501 and LC06014.  We thank all the developers of UK's AstroGrid Virtual
Observatory Project for testbeding  the new paradigm of scientific research
based on joining  supercomputing GRID technologies with Virtual Observatory
standards.  We are greatly indebted  to Pavel \v{S}koda and Jan Fuchs for
practical implementation of several versions of the KOREL Web Service.


\end{document}